# Effective-Medium Theory with No Approximation for One Dimensional Metamaterials

DONG WANG,[1] HOUZHI CAI,[1,*]

[1]*Key Laboratory of Optoelectronic Devices and Systems of Ministry of Education and Guangdong Province, College of Physics and Optoelectronic Engineering, Shenzhen University, Shenzhen 518060, China*

*\*hzcai@szu.edu.cn*

**Abstract:** Transfer matrix theory (TMT) is used to study the effective effective-medium theory (EMT) of one dimensional metamaterials (1D MMs). 1D MMs with the equal diagonal elements of periodic transfer matrix are defined as 1D perfect MMs (1D PMMs), which can be suitable for effective-medium theory (EMT) with no approximation. Then, two simplest types of 1D PMMs are derived. One type is the multilayer AB periodic structure, the simplest asymmetric structure consisting of two materials with equal impedance, i.e. $Z_a=Z_b$. The other type is the ABA periodic structure, the simplest symmetric structure without limitation. The normal incidence and oblique incidence characteristic of these two kinds of 1D PMM is studied. The expressions of effective refractive index, effective impedance, effective relative permeability, effective relative permittivity, and effective refractive angle are investigated and derived. It's interesting and confusing that the effective refractive angle doesn't obey the Snell's law, and is different from the effective geometrical refractive angle. Finite element method (FEM) and TMT are also used to further verify no approximation of our EMT. It is found that this kind of 1D PMM is an anisotropic material, and then the curves showing the changes of effective parameters with respect to incidence angle are also obtained.

## 1. Introduction

Optical metamaterials (MMs) have been intensively investigated for the last 20 years, because of their artificially structured with nanoscale inclusions and strikingly unconventional properties at optical frequencies [1-4]. MMs can be treated as macroscopically homogeneous media and can exhibit a variety of unusual and exciting responses to light, such as huge birefringence [5-7], negative refractive index [8-13], near-zero electric permittivity [14,15], electromagnetic wave cloaking [16], inverse Doppler Effect [17], and perfect absorber [18,19].

In modern photonics, there is the evident inclination in favor of numerical methods in order to describe, for example, the optical properties of metamaterials at the expense of physical intuition [20-23]. There is no doubt that modern numerical algorithms and available computer facilities provide the main way to investigate more or less complicated problems. Nevertheless, the qualitative approximate type models can provide a deeper understanding of the basic physical processes, stimulate discussion of new effects, and even provide new paradigm for optimization of a particular design. The qualitative models are complementary to the numerical ones, taking advantages of careful comparison with the results of rigorous numerical calculations, but at the same time remaining analytically treatable. In order to create this type of model, accurate approximations have to be made in order to simplify the respective consideration, at the same time keeping the main physical effects and interplay between them in the model, which is called the effective-medium theory (EMT) [24].

Originally, in 1973 EMT has been used to explain the Hall Effect in disordered materials by M. H. Cohen and J. Jortner [25], where they have presented a classical EMT for the magneto-conductivity tensor in disordered materials which undergo a metal-nonmetal transition via a microscopically inhomogeneous transport regime. They have defined an effective medium within which fluctuations of the field from the medium value average to zero. D. Stroud has generalized the EMT to treat inhomogeneous media with crystallites of arbitrary shape, size, and orientation, and conductivity tensors of arbitrary symmetry [26]. To obtain the generalization, they have first developed an integral equation for the electric field within the medium and then approximately decoupled the equation. The decoupling procedure serves to make clear the mean-field character of the EMT. D. E. Aspnes has implemented the definition of macroscopic quantities as averages of their microscopic counterparts [27]. This approach also leads naturally into a treatment of EMT and the description of the dielectric response of heterogeneous materials. K. W. Jacobsen et al have developed an extension of EMT from being able to description of all the atoms in a condensed system [28]. Based on EMT X. C. Zeng et al have proposed an approximate general method for calculating the effective dielectric function of a random composite in which there is a weakly nonlinear relation between electric displacement and electric field [29]. J. W. Haus et al have developed an EMT for nonlinear, ellipsoidal particles embedded in a host medium [30]. Specifically, they have treated spheroidal-shaped particles and give results for oriented and random configurations of the particles. They have also discussed metallic particles embedded in a linear dielectric medium and examined the enhancement of their nonlinear response at the surface-plasmon resonances.

As enter the 21st century, MMs have been introduced and rapidly been adopted as a means of achieving unique electromagnetic material response. In MMs, artificially structured, often periodically positioned, inclusions replace the atoms and molecules of conventional materials. The scale of these inclusions is smaller than that of the electromagnetic wavelength of interest, so that a homogenized description applies. Based on EMT D. R. Smith et al have analyzed the reflection and transmission coefficients calculated from transfer matrix simulations on finite lengths of MMs, to determine the effective permittivity and permeability [31]. Then, also based on EMT D. R. Smith and J. B. Pendry have presented a homogenization technique in which macroscopic fields are determined via averaging the local fields obtained from a full-wave electromagnetic simulation or analytical calculation [32]. The field-averaging method can be applied to homogenize any periodic structure with unit cells having inclusions of arbitrary geometry and material. To describe the effects of spatial dispersion of MMs, we need to generalize a standard homogenization theory introducing nonlocal effects. Such effects were observed in a number of different structures, and for some of them even an analytical description was suggested [33-41]. It worth noting that from first principles, an accurate homogenized description of MMs made of magneto dielectric inclusions has been derived to highlight and overcome relevant limitations of standard homogenization methods [42]. G. T. Papadakis et al have retrieved material parameters for uniaxial MMs [43]. S. Larouche et al have retrieved all effective susceptibilities in nonlinear MMs [44]. A. Rose et al have retrieved nonlinear parameter from three- and four-wave mixing in MMs [45].

These years one dimensional (1D) MMs have drawn much interesting for their simplest structure and analytic electromagnetic solutions (transfer matrix) [46,47]. A. V. Chebykin et al have built nonlocal effective medium model for multilayered metal-dielectric MMs [48]. M. A. Gorlach have studied the boundary conditions for the effective-medium description of 1D MMs [49]. However, the accumulation of qualitative models errors may result in the breakdown of effective-medium theory [50,51]. Then V. Popov et al have developed the operator approach to overcome this breakdown of Maxwell Garnett approximation [52].

However, all of above EMTs are developed due to some kinds of approximation. In this article, based on transfer matrix theory we develop an EMT with no approximation for 1D PMMs. We find that the transfer matrix of periodic unit with the equal diagonal elements can be replaced by an effective transfer matrix of effective homogenization material. However, the other kinds of periodic unit with the unequal diagonal elements cannot be replaced by any kinds of effective homogenization material. Therefore, we can define 1D MMs with the equal diagonal elements of transfer matrix as 1D perfect MMs (1D PMMs).

In order to prove the rigor of our theory, modern numerical algorithms, i.e. finite element method (FEM) and transfer matrix theory (TMT), have been used for comparison. COMSOL Multiphysics can be used for FEM analysis, and Essential Macleod can be used for TMT analysis.

In the following sections 2 and 3, the normal and the oblique incidence situations are studied, respectively.

## 2. Transfer Matrix for Normal Incidence

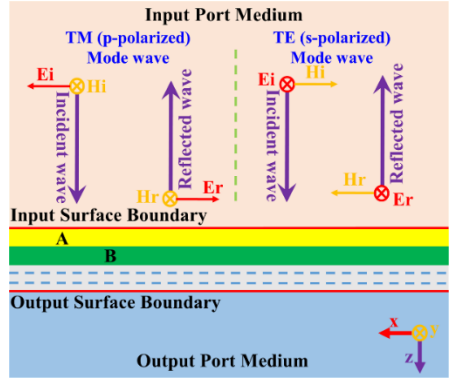

Fig. 1. Schematic diagram of multilayer structure with normal incidence by TM or TE mode wave.

First of all we define $\exp[i(\omega t - \vec{k}\cdot\vec{r})]$ as a wave propagating along the displacement vector $\vec{r}$. It means that the imaginary part of the refractive index is negative for absorbing media. In addition, as shown in Fig. 1, a wave with the electric vector in the plane of incidence is known as p-polarized or sometimes, as TM (for transverse magnetic) and a wave with the electric vector normal to the plane of incidence as s-polarized or, sometimes, TE (for transverse electric). Then we can obtain the transfer matrix of the $j$-th layer for normal incidence as (the detailed derivation is in section A of supplemental document)

$$T_j^{(p,s)} = \begin{bmatrix} T_{j(11)}^{(p,s)} & T_{j(12)}^{(p,s)} \\ T_{j(21)}^{(p,s)} & T_{j(22)}^{(p,s)} \end{bmatrix} = \begin{bmatrix} \cos(n_j k_0 d_j) & \pm i z_j \sin(n_j k_0 d_j) \\ \pm \dfrac{i}{z_j} \sin(n_j k_0 d_j) & \cos(n_j k_0 d_j) \end{bmatrix}, \quad (1)$$

where $k_0$ is the wave vector in vacuum, $z_j$ is the impedance ( $z_j = z_0\sqrt{\mu_{r(j)}/\varepsilon_{r(j)}}$, $\mu_{r(j)}$ and $\varepsilon_{r(j)}$ are the relative permeability and relative permittivity), $z_0$ is the impedance in vacuum ( $z_0 = \sqrt{\mu_0/\varepsilon_0}$, $\mu_0$ and $\varepsilon_0$ are the permeability and permittivity in vacuum), and $n_j$ is the refractive index( $n_j = \sqrt{\varepsilon_{r(j)}\mu_{r(j)}}$ ). However, if we define $\exp[-i(\omega t - \vec{k}\cdot\vec{r})]$ as a wave propagating along the displacement vector $\vec{r}$, the form of Eq. 1 is a little different, where the imaginary part of the refractive index is positive for absorbing media [31].

This result can be extended to the general case of assembly of $q$ layers, when the transfer matrix is simply the product of the individual matrices taken in the correct order, i.e.

$$\begin{bmatrix} E_{i\text{-}s}^{(p,s)} \\ H_{i\text{-}s}^{(p,s)} \end{bmatrix} = \left\{ \prod_{j=1}^{q} \begin{bmatrix} \cos(n_j k_0 d_j) & \pm i z_j \sin(n_j k_0 d_j) \\ \pm \frac{i}{z_j} \sin(n_j k_0 d_j) & \cos(n_j k_0 d_j) \end{bmatrix} \right\} \begin{bmatrix} E_{o\text{-}s}^{(p,s)} \\ H_{o\text{-}s}^{(p,s)} \end{bmatrix} = \begin{bmatrix} T_{11} & \pm T_{12} \\ \pm T_{21} & T_{22} \end{bmatrix} \begin{bmatrix} E_{o\text{-}s}^{(p,s)} \\ H_{o\text{-}s}^{(p,s)} \end{bmatrix}, \quad (2)$$

where the subscripts *i-s* and *o-s* means the physical quantities at the input surface boundary and the output surface boundary, respectively. The red lines in Fig. 1 indicates these two boundaries.

From Eq. (1), it can be found that the transfer matrix of a homogenization material must obey two rules. One is equal characteristic, i.e. that the diagonal elements of the transfer matrix should be equal. The other is unit characteristic, i.e. that the rank of the transfer matrix should be 1. After applying the EMT, the transfer matrix of periodic unit of any structure 1D MMs should obey these two rules. Otherwise, the 1D MMs cannot be treated as any kinds of effective homogenization material. This divides general 1D MMs into two categories. In this article, the 1D MMs obey these two rules are defined as 1D PMMs. Generally, the unit characteristic can be guaranteed for all 1D MMs. However, the equal characteristic is a little rigorous. Considering the symmetry, the configuration of 1D MMs can be divided into two kinds. One is asymmetric structure, and the other is symmetric structure. Therefore, AB periodic and ABA periodic structures are the simplest asymmetric and symmetric structures, respectively. In the following section 3 and section 4, we will study the normal incidence optical properties of AB periodic and ABA periodic structures. The optical properties are reflection and transmission coefficients. In detail, these coefficients are divided into amplitude reflection and transmission coefficients $\rho$ and $\tau$, and net irradiance reflection and transmission coefficients $R_I$ and $T_I$. In addition, the amplitude coefficient contains phase information.

From Eq. (2) amplitude reflection and transmission coefficients $\rho$ and $\tau$ can be derived as (the detailed derivation is in section B of supplemental document)

$$\rho^{(p,s)} = \frac{\mp z_{ou} T_{11} \mp T_{12} \pm z_{in} z_{ou} T_{21} \pm z_{in} T_{22}}{z_{ou} T_{11} + T_{12} + z_{in} z_{ou} T_{21} + z_{in} T_{22}}, \quad (3)$$

$$\tau^{(p,s)} = \frac{2 z_{ou}}{z_{ou} T_{11} + T_{12} + z_{in} z_{ou} T_{21} + z_{in} T_{22}}, \quad (4)$$

where $z_{in}$ and $z_{ou}$ are the impedances of the input port medium and the output port medium, respectively. The net irradiance $\vec{I}$ is defined as $\vec{I} = \frac{1}{2}\text{Re}(\vec{E} \times \vec{H}^*)$, and then the intense of net irradiance $I$ can be expressed as $I = \frac{1}{2}\text{Re}(EH^*) = \frac{1}{2}\text{Re}(EE^*/z^*) = \frac{1}{2}EE^* \text{Re}(1/z^*)$. The net irradiance reflection and transmission coefficients $R_I$ and $T_I$ can be expressed as [46,47]

$$R_I = I_{re}/I_{in} = \rho^{(p,s)}\left(\rho^{(p,s)}\right)^*, \quad (5)$$

$$T_I = I_{ou}/I_{in} = \tau^{(p,s)}\left(\tau^{(p,s)}\right)^* \text{Re}(1/z_{ou}^*)/\text{Re}(1/z_{in}^*). \quad (6)$$

Next, we will use the above equations to study the normal incidence properties of simplest AB and ABA periodic 1D MMs, and obtain the effective physical quantities. From Eqs. (1)-(6), it can be derived that: the impedance matching of the input port medium and single layer thin film medium will result in the net irradiance reflection not to change with the thickness of the thin film; the impedance matching of the output port medium and single layer thin film medium will result in the net irradiance transmission not to change with the thickness of the thin film. Not all 1D MMs can achieve impedance matching [50-52]. But for 1D PMMs, which satisfy the EMT with no approximation, the impedance matching is necessary and sufficient. Therefore, impedance matching is used as a way to check whether it is PMMs. Furthermore, FEM analysis, and TMT analysis are also used to prove the rigor of our theory for reference.

## 2.1. Normal Incidence for Simplest Asymmetric Structure

Firstly, we analyze the most studied AB periodic structure, which is the simplest asymmetric structure. For AB periodic structure the total transfer matrix of one period is

$$T^{AB(p,s)} = \begin{bmatrix} T_{11}^{AB} & T_{12}^{AB} \\ T_{21}^{AB} & T_{22}^{AB} \end{bmatrix}^{(p,s)} = \begin{bmatrix} T_{a(11)} & \pm T_{a(12)} \\ \pm T_{a(21)} & T_{a(22)} \end{bmatrix} \begin{bmatrix} T_{b(11)} & \pm T_{b(12)} \\ \pm T_{b(21)} & T_{b(22)} \end{bmatrix}$$
$$= \begin{bmatrix} T_{a(11)}T_{b(11)} + T_{a(12)}T_{b(21)} & \pm \left( T_{a(11)}T_{b(12)} + T_{a(12)}T_{b(22)} \right) \\ \pm \left( T_{a(21)}T_{b(11)} + T_{a(22)}T_{b(21)} \right) & T_{a(21)}T_{b(12)} + T_{a(22)}T_{b(22)} \end{bmatrix}. \quad (7)$$

Applying effective-medium theory, the matrix $T^{AB}$ must have the similar form with matrix $T_j$, as shown in Eq. (1). It's worth noting that the diagonal elements of matrix $T_j$ are equal, which is described as equal characteristic mentioned hereinbefore. However, the diagonal elements of matrix $T^{AB}$ are not equal, unless the impedances of the two materials A and B are equal to each other, i.e.

$$z_{a,b} = z_a = z_b. \quad (8)$$

This is a very harsh condition. It means that the two materials A and B should be impedance matched, i.e. $\mu_{r(a)}/\varepsilon_{r(a)} = \mu_{r(b)}/\varepsilon_{r(b)}$. However, it is not possible to match the impedance of all wavelengths. The impedance match can be achieved at the intersection point of the dispersion curves of the permittivity and permeability division.

Based on Eq. (8) and the effective-medium theory, the matrix $T^{AB}$ becomes the simple form as

$$T^{AB(p,s)} = \begin{bmatrix} \cos(n_a k_0 d_a + n_b k_0 d_b) & \pm i z_{a,b} \sin(n_a k_0 d_a + n_b k_0 d_b) \\ \pm \frac{i}{z_{a,b}} \sin(n_a k_0 d_a + n_b k_0 d_b) & \cos(n_a k_0 d_a + n_b k_0 d_b) \end{bmatrix}$$
$$= \begin{bmatrix} \cos\left[ n_{eff}^{AB} k_0 (d_a + d_b) \right] & \pm i z_{eff}^{AB} \sin\left[ n_{eff}^{AB} k_0 (d_a + d_b) \right] \\ \pm \frac{i}{z_{eff}^{AB}} \sin\left[ n_{eff}^{AB} k_0 (d_a + d_b) \right] & \cos\left[ n_{eff}^{AB} k_0 (d_a + d_b) \right] \end{bmatrix}, \quad (9)$$

$$z_{eff}^{AB} = z_{a,b}, \quad (10a)$$

$$n_{eff}^{AB} = \frac{n_a d_a + n_b d_b}{d_a + d_b}, \quad (10b)$$

$$\mu_{(r)eff}^{AB} = n_{eff}^{AB} z_{eff}^{AB} / z_0 = \frac{n_a d_a + n_b d_b}{d_a + d_b} \left( \frac{z_{a,b}}{z_0} \right), \quad (10c)$$

$$\varepsilon_{(r)eff}^{AB} = n_{eff}^{AB} z_0 / z_{eff}^{AB} = \frac{n_a d_a + n_b d_b}{d_a + d_b} \left( \frac{z_0}{z_{a,b}} \right), \quad (10d)$$

where $z_{eff}^{AB}$, $n_{eff}^{AB}$, $\mu_{(r)eff}^{AB}$, $\varepsilon_{(r)eff}^{AB}$ are, respectively, effective impedance, effective refractive index, effective relative permeability and effective relative permittivity of the AB periodic structure in the Eq. (8) condition.

As a concrete example, we take wavelength as 500 nm, $\mu_{r(a)} = 1.5$, $\varepsilon_{r(a)} = 3$, $\mu_{r(b)} = 1.2$, $\varepsilon_{r(b)} = 2.4$, and $d_a = d_b = 10 nm$. In this case, the two materials A and B are impedance matched, and so this AB periodic structure is a 1D PMM. From Eqs. (10a) and (10b), we can obtain the effective parameters $z_{eff}^{AB} = z_0/\sqrt{2}$, $n_{eff}^{AB} = 1.9092$, $\mu_{(r)eff}^{AB} = 1.3500$, and $\varepsilon_{(r)eff}^{AB} = 2.7000$. Next we study the electromagnetic wave transmission characteristics of this structure based on Eqs. (5) and (6), where only the effective impedance affects transmittance and reflectivity. Fig. 2(A) is about the output impedance matching, and Fig. 2(B) is about the input impedance matching. Concretely, we take the relative permeability of the input port medium $\mu_{r(in)}$ and output port medium $\mu_{r(ou)}$ as 1. Therefore, the relative permittivity to be 2 can achieve the impedance matching. In Fig. 2(A) the relative permittivity of the input port medium $\varepsilon_{r(in)}$ is 1, and the relative permittivity of the output port medium $\varepsilon_{r(ou)}$ varies from 1.7 to 2.3 by 0.1. In Fig. 2(B) the relative permittivity of the output port medium $\varepsilon_{r(ou)}$ is 5, and the relative permittivity of the input port medium $\varepsilon_{r(in)}$ varies from 1.7 to 2.3 by 0.1.

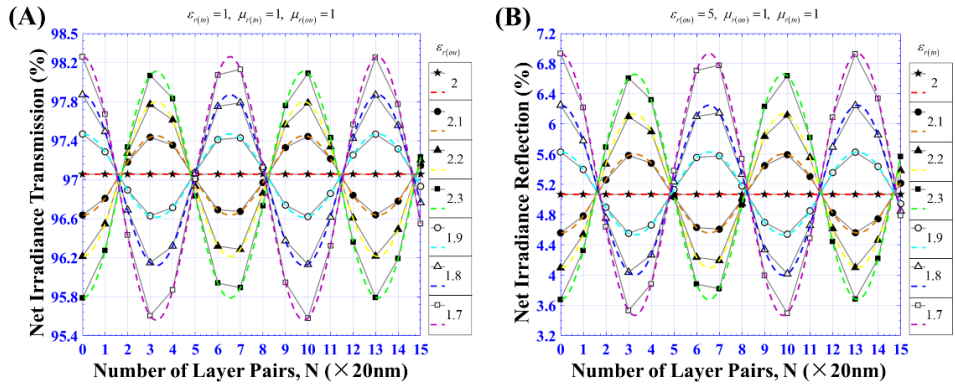

Fig. 2. Transmission characteristics of output (A) and input (B) impedance matching for AB periodic structure calculated by TMT. $\mu_{r(a)} = 1.5$, $\varepsilon_{r(a)} = 3$, $\mu_{r(b)} = 1.2$, $\varepsilon_{r(b)} = 2.4$, and $d_a = d_b = 10 nm$.

The dotted lines are the results by applying EMT and thinking of the AB periodic structure as a homogeneous material with the effective parameters using TMT. Essential Macleod can only be used for unit relative permeability situation, i.e. $\mu_r = 1$. Therefore, Matlab is used to set up the TMT physical model and then to obtain the dotted lines of EMT. The broke lines are the results of AB periodic structure with different numbers cycles, in which the endpoints are represented in various shapes. Solid pentacle, solid circle, solid triangle, and solid square are corresponding to parameters $\varepsilon_{r(ou)}$ (in Fig. 2(A)) and $\varepsilon_{r(in)}$ (in Fig. 2(B)) to be 2, 2.1, 2.2, and 2.3, respectively. Hollow circle, hollow triangle, and hollow square are corresponding to parameters $\varepsilon_{r(ou)}$ (in Fig. 2(A)) and $\varepsilon_{r(in)}$ (in Fig. 2(B)) to be 1.9, 1.8, and 1.7, respectively.

The results in Fig. 2 show that, when achieve impedance matching, i.e. input port impedance matching and output port impedance matching, the transmittance and reflectivity are horizontal lines that don't depend on the thickness. When impedance mismatch, the transmittance and reflectivity fluctuate with the thickness, and the fluctuation range is small enough with a little mismatch. For AB periodic structure, the thickness varies by number of AB period pair, and is discrete. For effective homogeneous material, the thickness varies continuously. The discrete shapes coincide perfectly with the doted curves. It means that there is no approximation between our EMT (i.e. Eqs. (9a)-(9d)) and TMT. That's why we call it perfect MMs. Essential Macleod based on TMT has been widely used to guide the research activities of design, analysis, and monitoring of optical film coatings [22,23,53-57]. TMT is the strict solution of Maxwell's equations for optical films [46,47].

To further verify no approximation of Eqs. (9a)-(9d), we use Comsol Multiphysics (a software based on FEM) to simulate our AB periodic structure and the corresponding EMT model. Figs. 3(A) and 3(B) are corresponding to Figs. 2(A) and 2(B), respectively. Because of the numerical error, the shape points and dotted lines calculated by FEM in Fig. 3 are a little different from those calculated by TMT in Fig. 2. Analogously, the shape points are the results of AB periodic structure with different numbers cycles, and the dotted lines are the results of EMT based on Eqs. (9a)-(9d). In Fig. 3 the discrete shapes also coincide perfectly with the doted curves, and it can also reflect that there are no approximation between our EMT (i.e. Eqs. (9a)-(9d)) and FEM.

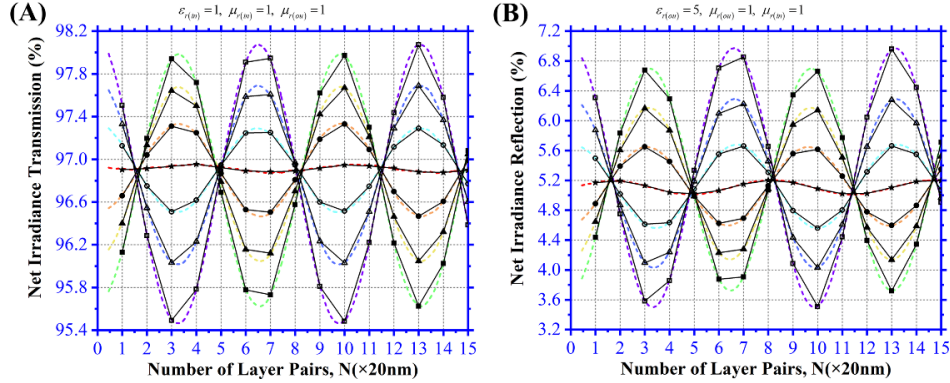

Fig. 3. Transmission characteristics of output (A) and input (B) impedance matching for AB periodic structure calculated by FEM. $\mu_{r(a)}=1.5$, $\varepsilon_{r(a)}=3$, $\mu_{r(b)}=1.2$, $\varepsilon_{r(b)}=2.4$, and $d_a=d_b=10nm$.

In the next sub-section we analyze the simplest symmetric structure, i.e. ABA periodic structure.

## 2.2. Normal Incidence for Simplest Symmetric Structure

In order to make sure the diagonal terms are equal, the simplest symmetric structure, i.e. ABA periodic structure is introduced. The ABA transfer matrix has the following formula

$$T^{ABA(p,s)} = \begin{bmatrix} \cos\left[n_{eff}^{ABA}k_0(d_a+d_b)\right] & \pm i z_{eff}^{ABA}\sin\left[n_{eff}^{ABA}k_0(d_a+d_b)\right] \\ \pm \frac{i}{z_{eff}^{ABA}}\sin\left[n_{eff}^{ABA}k_0(d_a+d_b)\right] & \cos\left[n_{eff}^{ABA}k_0(d_a+d_b)\right] \end{bmatrix} = \begin{bmatrix} T_{11}^{ABA} & T_{12}^{ABA} \\ T_{21}^{ABA} & T_{22}^{ABA} \end{bmatrix}^{(p,s)}$$

$$= \begin{bmatrix} T_{a(11)} & \pm T_{a(12)} \\ \pm T_{a(21)} & T_{a(22)} \end{bmatrix} \begin{bmatrix} T_{b(11)} & \pm T_{b(12)} \\ \pm T_{b(21)} & T_{b(22)} \end{bmatrix} \begin{bmatrix} T_{a(11)} & \pm T_{a(12)} \\ \pm T_{a(21)} & T_{a(22)} \end{bmatrix}. \tag{11}$$

$$T_{11}^{ABA} = T_{a(11)}T_{a(11)}T_{b(11)} + T_{a(11)}T_{a(12)}T_{b(21)} + T_{a(11)}T_{a(21)}T_{b(12)} + T_{a(12)}T_{a(21)}T_{b(22)}, \tag{12a}$$

$$T_{12}^{ABA} = \pm\left(T_{a(11)}T_{a(22)}T_{b(12)} + T_{a(11)}T_{a(12)}T_{b(11)} + T_{a(22)}T_{a(12)}T_{b(22)} + T_{a(12)}T_{a(12)}T_{b(21)}\right), \tag{12b}$$

$$T_{21}^{ABA} = \pm\left(T_{a(11)}T_{a(22)}T_{b(21)} + T_{a(11)}T_{a(21)}T_{b(11)} + T_{a(22)}T_{a(21)}T_{b(22)} + T_{a(21)}T_{a(21)}T_{b(12)}\right), \tag{12c}$$

$$T_{22}^{ABA} = T_{a(22)}T_{a(22)}T_{b(22)} + T_{a(12)}T_{a(22)}T_{b(21)} + T_{a(21)}T_{a(22)}T_{b(12)} + T_{a(12)}T_{a(21)}T_{b(11)}. \tag{12d}$$

In the case of equal diagonal elements for matrixes $T_a$ and $T_b$, the ABA matrix $T^{ABA}$ also has equal diagonal elements. Substitute into matrixes $T_a$ and $T_b$, effective impedance $z_{eff}^{ABA}$, effective refractive index $n_{eff}^{ABA}$, effective relative permeability $\mu_{(r)eff}^{ABA}$, and effective relative permittivity $\varepsilon_{(r)eff}^{ABA}$ of the ABA periodic structure can be written in the forms

$$n_{eff}^{ABA} = \frac{\arccos\left[T_{11}^{ABA(p,s)}\right]}{k_0(2d_a+d_b)}, \tag{13a}$$

$$z_{eff}^{ABA} = \sqrt{T_{12}^{ABA(p,s)}/T_{21}^{ABA(p,s)}}, \tag{13b}$$

$$\mu_{(r)eff}^{ABA} = n_{eff}^{ABA} z_{eff}^{ABA}/z_0, \tag{13c}$$

$$\varepsilon_{(r)eff}^{ABA} = n_{eff}^{ABA} z_0/z_{eff}^{ABA}. \tag{13d}$$

As a concrete example, we take wavelength also as 500 nm, $\mu_{r(a)}=1.5$, $\varepsilon_{r(a)}=5$, $\mu_{r(b)}=1$, $\varepsilon_{r(b)}=1$, and $d_a$=5 nm, $d_b$=10 nm, which are the same parameters as in reference [50].

In this ABA structure case, the two materials A and B have no similar limitation as the AB periodic structure. Specially, for the impedance match $z_a = z_b$ of ABA structure case, the effective impedance and effective refractive index are, respectively, $z_{eff} = z_a = z_b$, and $n_{eff} = (2n_a d_a + n_b d_b)/(2d_a + d_b)$. These effective medium formulas have some similarity with Eqs. (10a) and (10b).

From Eqs. (13a)-(13d), we can obtain the effective parameters $z_{eff}^{AB} = 0.5815Z_0$, $n_{eff}^{AB} = 1.7336$, $\mu_{(r)eff}^{ABA} = 1.0081$, and $\varepsilon_{(r)eff}^{ABA} = 2.9813$. Fig. 4(A) is about the output impedance matching, and Fig. 4(B) is about the input impedance matching. Concretely, we take the relative permeability of the input port medium $\mu_{r(in)}$ and output port medium $\mu_{r(ou)}$ as 1. Therefore, the relative permittivity to be 2.9572 can achieve the impedance matching. In Fig. 4(A) the relative permittivity of the input port medium $\varepsilon_{r(in)}$ is 1, and the relative permittivity of the output port medium $\varepsilon_{r(ou)}$ varies around 2.9572 by 0.1. In Fig. 4(B) the relative permittivity of the output port medium $\varepsilon_{r(ou)}$ is 5, and the relative permittivity of the input port medium $\varepsilon_{r(in)}$ varies around 2.9572 by 0.1.

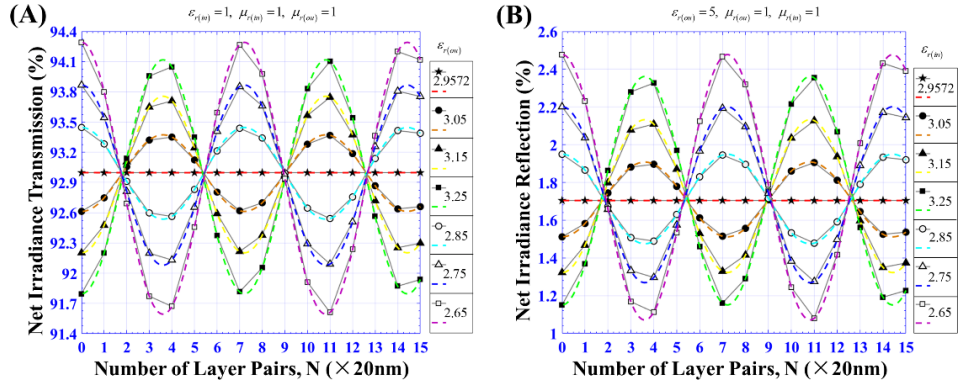

Fig. 4. Transmission characteristics of output (A) and input (B) impedance matching for ABA periodic structure calculated by TMT. $\mu_{r(a)}=1$, $\varepsilon_{r(a)}=5$, $\mu_{r(b)}=1$, $\varepsilon_{r(b)}=1$, and $d_a=5nm$, $d_b=10nm$.

The dotted lines are the results by applying EMT and thinking of the ABA periodic structure as a homogeneous material with the effective parameters. The broke lines are the results of ABA periodic structure with different numbers cycles, in which the endpoints are represented in various shapes. Solid pentacle, solid circle, solid triangle, and solid square are corresponding to parameters $\varepsilon_{r(ou)}$ (in Fig. 4(A)) and $\varepsilon_{r(in)}$ (in Fig. 4(B)) to be 2.9572, 3.05, 3.15, and 3.25, respectively. Hollow circle, hollow triangle, and hollow square are corresponding to parameters $\varepsilon_{r(ou)}$ (in Fig. 4(A)) and $\varepsilon_{r(in)}$ (in Fig. 4(B)) to be 2.85, 2.75, and 2.65, respectively.

The results in Fig. 4 show that, when achieve impedance matching, the transmittance and reflectivity are horizontal lines that don't depend on the thickness. When impedance mismatch, the transmittance and reflectivity fluctuate with the thickness, and the fluctuation range is small enough with a little mismatch. The discrete points also coincide perfectly with the doted curves. Therefore ABA periodic structure is a kind of 1D PMMs without any special limitation.

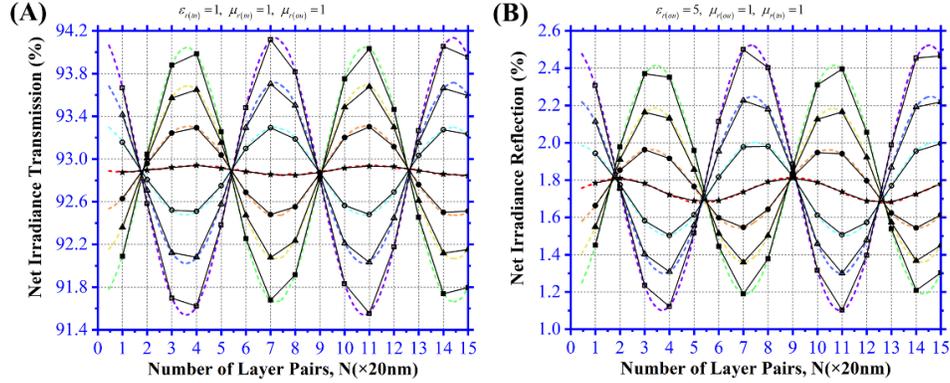

Fig. 5. Transmission characteristics of output (A) and input (B) impedance matching for ABA periodic structure calculated by FEM. $\mu_{r(a)}=1$, $\varepsilon_{r(a)}=5$, $\mu_{r(b)}=1$, $\varepsilon_{r(b)}=1$, and $d_a=5nm$, $d_b=10nm$.

To further verify no approximation of Eqs. (13a)-(13d), we also use Comsol Multiphysics to simulate our ABA periodic structure and the corresponding EMT model. Figs. 5(A) and 5(B) are corresponding to the Figs. 4(A) and 4(B), respectively. Because of the numerical error, the shape points and dotted lines calculated by FEM in Fig. 5 are a little different from those calculated by TMT in Fig. 4. Analogously, the shape points are the results of ABA periodic structure with different numbers cycles, and the dotted lines are the results of EMT based on Eqs. (13a)-(13d). In Fig. 5 the discrete shapes also coincide perfectly with the doted curves, and it can also reflect that there is no approximation between our EMT (i.e. Eqs. (13a)-(13d)) and FEM.

## 3. Transfer Matrix for Oblique Incidence

In this section, oblique incidence characteristic of these two kinds of 1D PMM, i.e. AB and ABA structure, is studied. Different from normal incidence, the effective physical parameters of oblique incidence is related to incidence angle $\theta_{in}$, as shown in Fig. 6. We define $\exp[i(\omega t - \vec{k}\cdot\vec{r})]$ as a wave propagating along the displacement vector $\vec{r}$.

First of all, the transfer matrix of the $j$-th layer for oblique incidence can be obtained as (the detailed derivation is in section C of supplemental document)

$$T_j^{p,s}\Phi_j = \begin{bmatrix} T_{j(11)}^{p,s} & T_{j(12)}^{p,s} \\ T_{j(21)}^{p,s} & T_{j(22)}^{p,s} \end{bmatrix}\Phi_j$$

$$= \begin{bmatrix} \cos(n_j k_0 d_j \cos\theta_j) & \pm i z_j^{p,s}\sin(n_j k_0 d_j \cos\theta_j) \\ \pm \dfrac{i}{z_j^{p,s}}\sin(n_j k_0 d_j \cos\theta_j) & \cos(n_j k_0 d_j \cos\theta_j) \end{bmatrix}\exp(-i n_j k_0 \sin\theta_j \tan\theta_j d_j), \quad (14a)$$

$$z_j^{(t)p} = z_j\cos\theta_j, \tag{14b}$$

$$z_j^{(t)s} = \frac{z_j}{\cos\theta_j}, \tag{14c}$$

where $k_0$ is the wave vector in vacuum, $\theta_j$ is the refraction angle, $z_j$ is the impedance ($z_j = z_0\sqrt{\mu_{r(j)}/\varepsilon_{r(j)}}$), $z_0$ is the impedance in vacuum ($z_0 = \sqrt{\mu_0/\varepsilon_0}$), and $n_j$ is the refractive index ($n_j = \sqrt{\varepsilon_{r(j)}\mu_{r(j)}}$). In addition, $z_j^{(t)p}$ and $z_j^{(t)s}$ are the tilted impedances for p- and s-polarized waves, respectively. It's noting that $\Phi_j$ is a phase factor due to the wave propagating along the $x$ direction for oblique incidence situation. Snell's law gives another relationship between the refraction angle $\theta_j$ and incidence angle $\theta_{in}$, as $\text{Re}(n_j)\sin\theta_j = \text{Re}(n_{in})\sin\theta_{in}$.

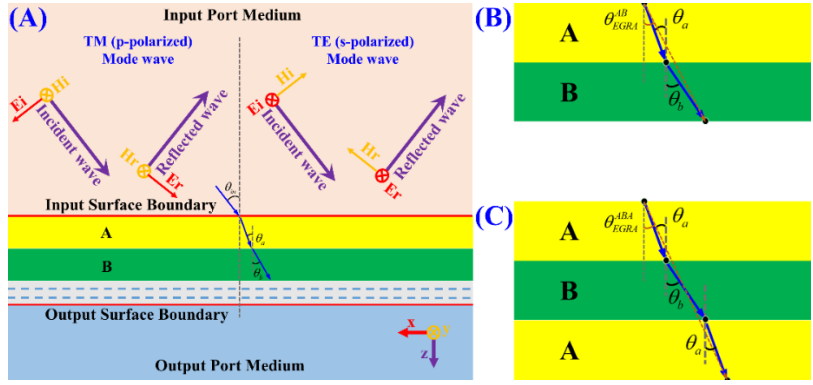

Fig. 6. (A) Schematic diagram of multilayer structure with oblique incidence by TM or TE mode wave. (B) Schematic diagram of EGRA for AB periodic structure. (C) Schematic diagram of EGRA for ABA periodic structure.

Referring to Eq. (2), we can get a similar formula as follows

$$\begin{bmatrix} E_{i\text{-}s}^{\parallel} \\ H_{i\text{-}s}^{\parallel} \end{bmatrix} = \left\{ \prod_{j=1}^{q} \begin{bmatrix} \cos(n_j k_0 d_j \cos\theta_j) & \pm i z_j^{(t)(p,s)} \sin(n_j k_0 d_j \cos\theta_j) \\ \pm \dfrac{i}{z_j^{(t)(p,s)}} \sin(n_j k_0 d_j \cos\theta_j) & \cos(n_j k_0 d_j \cos\theta_j) \end{bmatrix} \Phi_j \right\} \begin{bmatrix} E_{o\text{-}s}^{\parallel} \\ H_{o\text{-}s}^{\parallel} \end{bmatrix}$$

$$= \begin{bmatrix} T_{11}^{(p,s)} & \pm T_{12}^{(p,s)} \\ \pm T_{21}^{(p,s)} & T_{22}^{(p,s)} \end{bmatrix} \exp\left[ -ik_0 \sum_{j=1}^{q} (n_j \sin\theta_j \tan\theta_j d_j) \right] \begin{bmatrix} E_{o\text{-}s}^{\parallel} \\ H_{o\text{-}s}^{\parallel} \end{bmatrix}, \tag{15}$$

where $E_{i\text{-}s}^{\parallel}$, $H_{i\text{-}s}^{\parallel}$, $E_{o\text{-}s}^{\parallel}$ and $H_{o\text{-}s}^{\parallel}$ are the paralell components of $E_{i\text{-}s}$, $H_{i\text{-}s}$, $E_{o\text{-}s}$ and $H_{o\text{-}s}$, respectively, i.e. the components parallel to the boundary, which are connected by the tilted impedances $z_j^{(t)p}$ and $z_j^{(t)s}$. The subscripts $i$-$s$ and $o$-$s$ means the physical quantities at the input surface boundary and the output surface boundary, respectively. Then, from Eq. (15) amplitude reflection and transmission coefficients $\rho$ and $\tau$ can be derived as

$$\rho^{\parallel(p,s)} = \frac{\mp z_{ou}^{(t)(p,s)} T_{11}^{(p,s)} \mp T_{12}^{(p,s)} \pm z_{in}^{(t)(p,s)} z_{ou}^{(t)(p,s)} T_{21}^{(p,s)} \pm z_{in}^{(t)(p,s)} T_{22}^{(p,s)}}{z_{ou}^{(t)(p,s)} T_{11}^{(p,s)} + T_{12}^{(p,s)} + z_{in}^{(t)(p,s)} z_{ou}^{(t)(p,s)} T_{21}^{(p,s)} + z_{in}^{(t)(p,s)} T_{22}^{(p,s)}}, \tag{16}$$

$$\tau^{\parallel(p,s)} = \frac{2 z_{ou}^{(t)(p,s)} \exp\left[ ik_0 \sum_{j=1}^{q} (n_j \sin\theta_j \tan\theta_j d_j) \right]}{z_{ou}^{(t)(p,s)} T_{11}^{(p,s)} + T_{12}^{(p,s)} + z_{in}^{(t)(p,s)} z_{ou}^{(t)(p,s)} T_{21}^{(p,s)} + z_{in}^{(t)(p,s)} T_{22}^{(p,s)}}, \tag{17}$$

where $z_{in}^{(t)p}$ and $z_{in}^{(t)s}$ are the tilted impedances of the incoming port medium, $z_{ou}^{(t)p}$ and $z_{ou}^{(t)s}$ are the tilted impedances of the output port medium, and $\theta_{ou}$ are the refraction angle on the output area. The net irradiance reflection and transmission coefficients $R_I$ and $T_I$ can be expressed as

$$R_I = \frac{I_{re}}{I_{in}} = \rho^{\|(p,s)}\left(\rho^{\|(p,s)}\right)^*, \tag{18}$$

$$T_I = \frac{I_{ou}}{I_{in}} = \tau^{\|(p,s)}\left(\tau^{\|(p,s)}\right)^* \frac{\mathrm{Re}(1/z_{ou}^{(t)(p,s)})}{\mathrm{Re}(1/z_{in}^{(t)(p,s)})}. \tag{19}$$

Next, we will use the above equations to study the two kinds of 1D PMMs.

### 3.1. Oblique Incidence for Simplest Asymmetric Structure

Firstly, we analyze the most studied AB periodic structure. For AB periodic structure the total transfer matrix of one period is

$$T^{AB}\Phi^{AB} = \begin{bmatrix} T_{11}^{AB} & T_{12}^{AB} \\ T_{21}^{AB} & T_{22}^{AB} \end{bmatrix}\Phi^{AB} = \begin{bmatrix} T_{a(11)} & T_{a(12)}^{p,s} \\ T_{a(21)}^{p,s} & T_{a(22)} \end{bmatrix}\begin{bmatrix} T_{b(11)} & T_{b(12)}^{p,s} \\ T_{b(21)}^{p,s} & T_{b(22)} \end{bmatrix}\Phi^{AB}$$

$$= \begin{bmatrix} T_{a(11)}T_{b(11)} + T_{a(12)}^{p,s}T_{b(21)}^{p,s} & T_{a(11)}T_{b(12)}^{p,s} + T_{a(12)}^{p,s}T_{b(22)} \\ T_{a(21)}^{p,s}T_{b(11)} + T_{a(22)}T_{b(21)}^{p,s} & T_{a(21)}^{p,s}T_{b(12)}^{p,s} + T_{a(22)}T_{b(22)} \end{bmatrix}\Phi^{AB}, \tag{20}$$

$$T_{11}^{AB} = \left[\cos(n_a k_0 d_a \cos\theta_a)\cos(n_b k_0 d_b \cos\theta_b) - \frac{z_a^{(t)(p,s)}}{z_b^{(t)(p,s)}}\sin(n_a k_0 d_a \cos\theta_a)\sin(n_b k_0 d_b \cos\theta_b)\right], \tag{21a}$$

$$T_{12}^{AB} = \left[iz_b^{(t)(p,s)}\cos(n_a k_0 d_a \cos\theta_a)\sin(n_b k_0 d_b \cos\theta_b) + iz_a^{(t)(p,s)}\sin(n_a k_0 d_a \cos\theta_a)\cos(n_b k_0 d_b \cos\theta_b)\right], \tag{21b}$$

$$T_{21}^{AB} = \left[\frac{i}{z_a^{(t)(p,s)}}\sin(n_a k_0 d_a \cos\theta_a)\cos(n_b k_0 d_b \cos\theta_b) + \frac{i}{z_b^{(t)(p,s)}}\cos(n_a k_0 d_a \cos\theta_a)\sin(n_b k_0 d_b \cos\theta_b)\right], \tag{21c}$$

$$T_{22}^{AB} = \left[\cos(n_a k_0 d_a \cos\theta_a)\cos(n_b k_0 d_b \cos\theta_b) - \frac{z_b^{(t)(p,s)}}{z_a^{(t)(p,s)}}\sin(n_a k_0 d_a \cos\theta_a)\sin(n_b k_0 d_b \cos\theta_b)\right]. \tag{21d}$$

$$\Phi^{AB} = \exp\left[-ik_0\left(n_a \sin\theta_a \tan\theta_a d_a + n_b \sin\theta_b \tan\theta_b d_b\right)\right]. \tag{21e}$$

Applying effective-medium theory, the AB periodic matrix $T^{AB}\Phi^{AB}$ must have the similar form with transfer matrix $T_j^{p,s}\Phi_j$. It's worth noting that the diagonal elements of transfer matrix $T_j^{p,s}$ are equal. However the diagonal elements of AB periodic matrix $T^{AB}$ are not equal, unless the p- and s-polarized impedances of the two materials A and B are equal, i.e.

$$z_{a,b}^{(t)(p,s)} = z_a^{(t)(p,s)} = z_b^{(t)(p,s)}. \tag{22}$$

This is a very harsh condition. Based on Eq. (22), the matrix $T^{AB}\Phi^{AB}$ becomes the simple form as

$$T^{AB}\Phi^{AB} = T_{eff}^{AB}\Phi_{eff}^{AB}, \tag{23a}$$

$$T_{eff(11)}^{AB} = T_{eff(22)}^{AB} = \cos\left[n_{eff}^{AB} k_0 (d_a + d_b)\cos\theta_{eff}^{AB}\right], \tag{23b}$$

$$T_{eff(12)}^{AB} = iz_{eff}^{AB(t)(p,s)}\sin\left[n_{eff}^{AB} k_0 (d_a + d_b)\cos\theta_{eff}^{AB}\right], \tag{23c}$$

$$T_{eff(21)}^{AB} = \frac{i}{z_{a,b}^{(t)(p,s)}} \sin(n_a k_0 d_a \cos\theta_a + n_b k_0 d_b \cos\theta_b), \tag{23d}$$

$$\Phi_{eff}^{AB} = \exp\left[-ik_0 n_{eff}^{AB} \sin\theta_{eff}^{AB} \tan\theta_{eff}^{AB}(d_a + d_b)\right], \tag{23e}$$

where $z_{eff}^{AB(t)(p,s)}$, $n_{eff}^{AB}$, and $\theta_{eff}^{AB}$ are, respectively, effective tilted impedance, effective refractive index, and effective refractive angle of the AB periodic structure in the Eq. (22) condition. Based on Eqs. (23a)-(23e), we can obtain the following equations

$$n_{eff}^{AB}(d_a + d_b)\cos\theta_{eff}^{AB} = n_a d_a \cos\theta_a + n_b d_b \cos\theta_b, \tag{24a}$$

$$n_{eff}^{AB}\sin\theta_{eff}^{AB}\tan\theta_{eff}^{AB}(d_a + d_b) = n_a \sin\theta_a \tan\theta_a d_a + n_b \sin\theta_b \tan\theta_b d_b. \tag{24b}$$

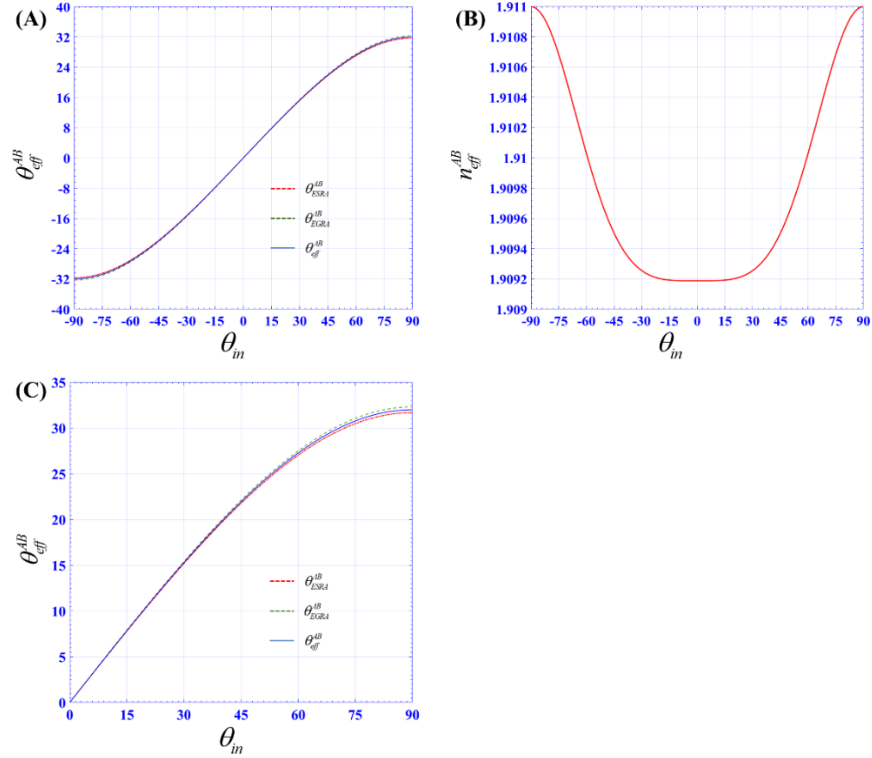

Fig. 7. (A) Effective refractive angle $\theta_{eff}^{AB}$ (solid blue line), effective Snell refractive angle $\theta_{ESRA}^{AB}$ (dotted red line) and effective geometrical refractive angle $\theta_{EGRA}^{AB}$ (dotted green line) with respect to incidence angle $\theta_{in}$. (B) Effective refractive index $n_{eff}^{AB}$ with respect to incidence angle $\theta_{in}$. (C) Partial enlarged figure of (A). $\mu_{r(a)} = 1.5$, $\varepsilon_{r(a)} = 3$, $\mu_{r(b)} = 1.2$, $\varepsilon_{r(b)} = 2.4$, and $d_a = d_b = 10nm$.

Then the effective refractive angle can be derived as

$$\tan\theta_{eff}^{AB} = \sqrt{\frac{n_a \sin\theta_a \tan\theta_a d_a + n_b \sin\theta_b \tan\theta_b d_b}{n_a d_a \cos\theta_a + n_b d_b \cos\theta_b}}, \quad (\theta_{in} \geq 0) \tag{25a}$$

$$\tan\theta_{eff}^{AB} = -\sqrt{\frac{n_a \sin\theta_a \tan\theta_a d_a + n_b \sin\theta_b \tan\theta_b d_b}{n_a d_a \cos\theta_a + n_b d_b \cos\theta_b}}. \quad (\theta_{in} < 0) \tag{25b}$$

We assume that the effective medium also obeys the Snell's law

$$\mathrm{Re}\left(n_{eff}^{AB}\right)\sin\theta_{eff}^{AB} = \mathrm{Re}(n_a)\sin\theta_a = \mathrm{Re}(n_b)\sin\theta_b. \tag{26}$$

Therefore, based on Eqs. (23) and (24a) the effective Snell refractive angle (ESRA) can be obtained as

$$\tan\theta_{ESRA}^{AB} = \frac{\sin\theta_a \sin\theta_b (d_a+d_b)}{\cos\theta_a \sin\theta_b d_a + \sin\theta_a \cos\theta_b d_b}. \tag{27}$$

However, the effective geometrical refractive angle (EGRA) can be derived by geometrical optics in one cycle of AB periodic structure, as shown in Fig. 6(B)

$$\tan\theta_{EGRA}^{AB} = \frac{\sin\theta_a \cos\theta_b d_a + \cos\theta_a \sin\theta_b d_b}{\cos\theta_a \cos\theta_b (d_a+d_b)}. \tag{28}$$

Obviously, effective refractive angle $\theta_{eff}^{AB}$ is different from ESRA $\theta_{ESRA}^{AB}$ and EGRA $\theta_{EGRA}^{AB}$. Physically, EGRA $\theta_{EGRA}^{AB}$ can reflect the path that the light actually traveled. So here's the interesting conclusion that the effective refractive angle $\theta_{eff}^{AB}$ doesn't obey the Snell's law and is different from the light path.

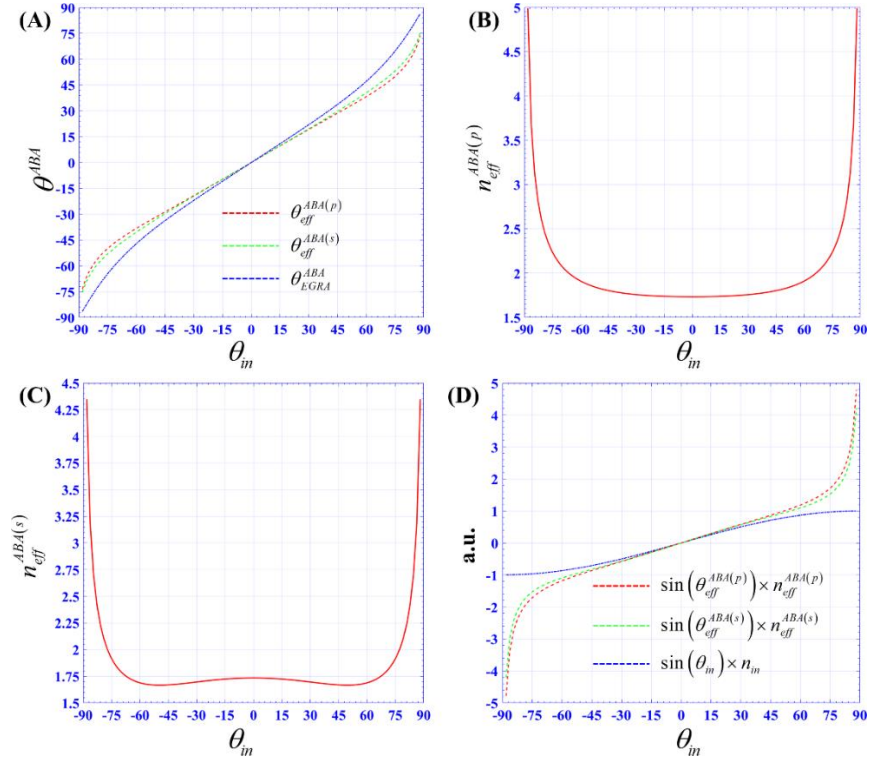

Fig. 8. (A) Effective refractive angele $\theta_{eff}^{ABA(p,s)}$ and effective geometrical refractive angle $\theta_{EGRA}^{AB}$ with respect to incidence angle $\theta_{in}$. (B) P-polarized effective refractive index $n_{eff}^{ABA(p)}$ with respect to incidence angle $\theta_{in}$. (C) S-polarized effective refractive index $n_{eff}^{ABA(s)}$ with respect to incidence angle $\theta_{in}$. (D) Breakdown of the Snell's law for effective-medium theory. $\mu_{r(a)}=1$, $\varepsilon_{r(a)}=5$, $\mu_{r(b)}=1$, $\varepsilon_{r(b)}=1$, and $d_a=5nm$, $d_b=10nm$.

Based on Eqs. (23) and (24), effective refractive index $n_{eff}^{AB}$, effective impedance $z_{eff}^{AB}$, effective tilted impedance $z_{eff}^{AB(t)(p,s)}$, effective relative permeability $\mu_{(r)eff}^{AB}$, and effective relative permittivity $\varepsilon_{(r)eff}^{AB}$ of the AB periodic structure with the incidence angle $\theta$ can be written in the forms

$$n_{eff}^{AB(\theta)} = \frac{\sqrt{(n_a\cos\theta_a d_a + n_b\cos\theta_b d_b)(n_a\cos\theta_a d_a + n_b\cos\theta_b d_b + n_a\sin\theta_a\tan\theta_a d_a + n_b\sin\theta_b\tan\theta_b d_b)}}{d_a + d_b}, \quad (29a)$$

$$z_{eff}^{AB(t)(p,s)} = z_{a,b}^{(t)(p,s)}, \quad (29b)$$

$$z_{eff}^{AB} = z_{a,b}, \quad (29c)$$

$$\mu_{(r)eff}^{AB(\theta)} = n_{eff}^{AB(\theta)} z_{eff}^{AB} / z_0, \quad (29d)$$

$$\varepsilon_{(r)eff}^{AB(\theta)} = n_{eff}^{AB(\theta)} z_0 / z_{eff}^{AB}. \quad (29e)$$

As the same example in section 2.1, we take wavelength as 500 nm, $\mu_{r(a)} = 1.5$, $\varepsilon_{r(a)} = 3$, $\mu_{r(b)} = 1.2$, $\varepsilon_{r(b)} = 2.4$, and $d_a = d_b = 10 nm$. In this case, the two materials A and B are impedance matched, and so this AB periodic structure is a 1D PMM. From Eqs. (25), (27), (28) and (29a), we can obtain the effective parameters $\theta_{eff}^{AB}$ $\theta_{ESRA}^{AB}$, $\theta_{EGRA}^{AB}$, and $n_{eff}^{AB}$ with respect to incidence angle $\theta_{in}$. In Fig. 7(A), the solid blue line ($\theta_{eff}^{AB}$) is a little different from dotted green line ($\theta_{EGRA}^{AB}$) and the dotted red line ($\theta_{ESRA}^{AB}$). In Fig. 7(B), the effective refractive index $n_{eff}^{AB}$ varies above 1.9092 by a small change.

### 3.2. Oblique Incidence for Simplest Symmetric Structure

Finally, we analyze the ABA periodic structure. For ABA periodic structure the total transfer matrix of one period is

$$T^{ABA}\Phi^{ABA} = \begin{bmatrix} T_{11}^{ABA} & T_{12}^{ABA} \\ T_{21}^{ABA} & T_{22}^{ABA} \end{bmatrix}^{(p,s)} \Phi^{ABA} = \begin{bmatrix} T_{a(11)} & T_{a(12)}^{p,s} \\ T_{a(21)}^{p,s} & T_{a(22)} \end{bmatrix}\begin{bmatrix} T_{b(11)} & T_{b(12)}^{p,s} \\ T_{b(21)}^{p,s} & T_{b(22)} \end{bmatrix}\begin{bmatrix} T_{a(11)} & T_{a(12)}^{p,s} \\ T_{a(21)}^{p,s} & T_{a(22)} \end{bmatrix}\Phi^{ABA}, \quad (30a)$$

$$T_{11}^{ABA(p,s)} = T_{a(11)}T_{a(11)}T_{b(11)} + T_{a(11)}T_{a(12)}^{p,s}T_{b(21)}^{p,s} + T_{a(11)}T_{a(21)}^{p,s}T_{b(12)}^{p,s} + T_{a(12)}^{p,s}T_{a(21)}^{p,s}T_{b(22)}, \quad (30b)$$

$$T_{12}^{ABA(p,s)} = T_{a(11)}T_{a(22)}T_{b(12)}^{p,s} + T_{a(11)}T_{a(12)}^{p,s}T_{b(11)} + T_{a(22)}T_{a(12)}^{p,s}T_{b(22)} + T_{a(12)}^{p,s}T_{a(12)}^{p,s}T_{b(21)}^{p,s}, \quad (30c)$$

$$T_{21}^{ABA(p,s)} = T_{a(11)}T_{a(22)}T_{b(21)}^{p,s} + T_{a(11)}T_{a(21)}^{p,s}T_{b(11)} + T_{a(22)}T_{a(21)}^{p,s}T_{b(22)} + T_{a(21)}^{p,s}T_{a(21)}^{p,s}T_{b(12)}^{p,s}, \quad (30d)$$

$$T_{22}^{ABA(p,s)} = T_{a(22)}T_{a(22)}T_{b(22)} + T_{a(12)}^{p,s}T_{a(22)}T_{b(21)}^{p,s} + T_{a(21)}^{p,s}T_{a(22)}T_{b(12)}^{p,s} + T_{a(12)}^{p,s}T_{a(21)}^{p,s}T_{b(11)}, \quad (30e)$$

$$\Phi^{ABA} = \Phi^A \Phi^B \Phi^A, \quad (30f)$$

$$\begin{aligned}T_{11}^{ABA(p,s)} &= \cos^2(n_a k_0 d_a \cos\theta_a)\cos(n_b k_0 d_b \cos\theta_b) - \sin^2(n_a k_0 d_a \cos\theta_a)\cos(n_b k_0 d_b \cos\theta_b) \\ &\quad -\left(\frac{z_a^{(t)(p,s)}}{z_b^{(t)(p,s)}} + \frac{z_b^{(t)(p,s)}}{z_a^{(t)(p,s)}}\right)\sin(n_a k_0 d_a \cos\theta_a)\cos(n_a k_0 d_a \cos\theta_a)\sin(n_b k_0 d_b \cos\theta_b) \\ &= \cos\left[n_{eff}^{ABA(p,s)} k_0 (2d_a + d_b)\cos\theta_{eff}^{ABA(p,s)}\right],\end{aligned} \quad (31a)$$

$$T_{12}^{ABA(p,s)} = 2iz_a^{(t)(p,s)} \sin(n_a k_0 d_a \cos\theta_a)\cos(n_a k_0 d_a \cos\theta_a)\cos(n_b k_0 d_b \cos\theta_b)$$

$$+iz_b^{(t)(p,s)} \cos^2(n_a k_0 d_a \cos\theta_a)\sin(n_b k_0 d_b \cos\theta_b) - i\frac{\left(z_a^{(t)(p,s)}\right)^2}{z_b^{(t)(p,s)}}\sin^2(n_a k_0 d_a \cos\theta_a)\sin(n_b k_0 d_b \cos\theta_b) \quad (31b)$$

$$= iz_{eff}^{ABA(t)(p,s)} \sin\left[n_{eff}^{ABA(p,s)} k_0 (2d_a + d_b)\cos\theta_{eff}^{ABA(p,s)}\right],$$

$$T_{21}^{ABA(p,s)} = \frac{2i}{z_a^{(t)(p,s)}} \sin(n_a k_0 d_a \cos\theta_a)\cos(n_a k_0 d_a \cos\theta_a)\cos(n_b k_0 d_b \cos\theta_b)$$

$$+\frac{i}{z_b^{(t)(p,s)}} \cos^2(n_a k_0 d_a \cos\theta_a)\sin(n_b k_0 d_b \cos\theta_b) - i\frac{z_b^{(t)(p,s)}}{\left(z_a^{(t)(p,s)}\right)^2}\sin^2(n_a k_0 d_a \cos\theta_a)\sin(n_b k_0 d_b \cos\theta_b) \quad (31c)$$

$$= \frac{i}{z_{eff}^{ABA(t)(p,s)}} \sin\left[n_{eff}^{ABA(p,s)} k_0 (2d_a + d_b)\cos\theta_{eff}^{ABA(p,s)}\right],$$

$$T_{22}^{ABA(p,s)} = \cos^2(n_a k_0 d_a \cos\theta_a)\cos(n_b k_0 d_b \cos\theta_b) - \sin^2(n_a k_0 d_a \cos\theta_a)\cos(n_b k_0 d_b \cos\theta_b)$$

$$-\left(\frac{z_a^{(t)(p,s)}}{z_b^{(t)(p,s)}} + \frac{z_b^{(t)(p,s)}}{z_a^{(t)(p,s)}}\right)\sin(n_a k_0 d_a \cos\theta_a)\cos(n_a k_0 d_a \cos\theta_a)\sin(n_b k_0 d_b \cos\theta_b) \quad (31d)$$

$$= \cos\left[n_{eff}^{ABA(p,s)} k_0 (2d_a + d_b)\cos\theta_{eff}^{ABA(p,s)}\right],$$

$$\Phi^{ABA} = \exp\left[-ik_0(2n_a \sin\theta_a \tan\theta_a d_a + n_b \sin\theta_b \tan\theta_b d_b)\right]$$
$$= \exp\left[-ik_0 n_{eff} \sin\theta_{eff} \tan\theta_{eff} (2d_a + d_b)\right], \quad (31e)$$

where $z_{eff}^{ABA(t)(p,s)}$, $n_{eff}^{ABA}$, and $\theta_{eff}^{ABA}$ are, respectively, effective tilted impedance, effective refractive index, and effective refractive angle of the AB periodic structure. Based on Eqs. (30) and (31), we can obtain the following equations

$$n_{eff}^{AB}(d_a + d_b)\cos\theta_{eff}^{AB} = \arccos\left(T_{11}^{ABA(p,s)}\right), \quad (32a)$$

$$n_{eff}^{AB} \sin\theta_{eff}^{AB} \tan\theta_{eff}^{AB} (2d_a + d_b) = 2n_a \sin\theta_a \tan\theta_a d_a + n_b \sin\theta_b \tan\theta_b d_b. \quad (32b)$$

Then the effective refractive angle can be derived as

$$\tan\theta_{eff}^{ABA(p,s)} = \sqrt{\frac{k_0(2n_a \sin\theta_a \tan\theta_a d_a + n_b \sin\theta_b \tan\theta_b d_b)}{\arccos\left(T_{11}^{ABA(p,s)}\right)}}, \quad (\theta_{in} \geq 0) \quad (33a)$$

$$\tan\theta_{eff}^{ABA(p,s)} = -\sqrt{\frac{k_0(2n_a \sin\theta_a \tan\theta_a d_a + n_b \sin\theta_b \tan\theta_b d_b)}{\arccos\left(T_{11}^{ABA(p,s)}\right)}}. \quad (\theta_{in} < 0) \quad (33b)$$

The EGRA $\theta_{EGRA}^{ABA}$ can also be derived by geometrical optics in one cycle of ABA periodic structure, as shown in Fig. 6(C)

$$\tan\theta_{EGRA}^{ABA} = \frac{2\sin\theta_a \cos\theta_b d_a + \cos\theta_a \sin\theta_b d_b}{\cos\theta_a \cos\theta_b (2d_a + d_b)}. \quad (34)$$

Based on Eqs. (30)-(33), effective refractive index $n_{eff}^{ABA}$, effective impedance $z_{eff}^{ABA}$, effective tilted impedance $z_{eff}^{ABA(t)(p,s)}$, effective relative permeability $\mu_{(r)eff}^{ABA}$, and effective relative permittivity $\varepsilon_{(r)eff}^{ABA}$ of the ABA periodic structure with the incidence angle $\theta$ can be written in the forms

$$n_{eff}^{ABA(\theta)(p,s)} = \frac{\arccos\left(T_{11}^{ABA(p,s)}\right)}{k_0(2d_a + d_b)\cos\theta_{eff}^{ABA(p,s)}}, \tag{35a}$$

$$z_{eff}^{ABA(\theta)(t)(p,s)} = \sqrt{T_{12}^{ABA}/T_{21}^{ABA}}, \tag{35b}$$

$$z_{eff}^{ABA(\theta)(p)} = z_{eff}^{ABA(\theta)(t)(p)}/\cos\theta_{eff}^{ABA(p,s)}, \tag{35c}$$

$$z_{eff}^{ABA(\theta)(s)} = z_{eff}^{ABA(\theta)(t)(s)}\cos\theta_{eff}^{ABA(p,s)}, \tag{35d}$$

$$\mu_{(r)eff}^{ABA(\theta)(p,s)} = n_{eff}^{ABA(\theta)(p,s)}z_{eff}^{ABA(\theta)(p,s)}/z_0, \tag{35e}$$

$$\varepsilon_{(r)eff}^{ABA(\theta)(p,s)} = n_{eff}^{ABA(\theta)(p,s)}z_0/z_{eff}^{ABA(\theta)(p,s)}. \tag{35f}$$

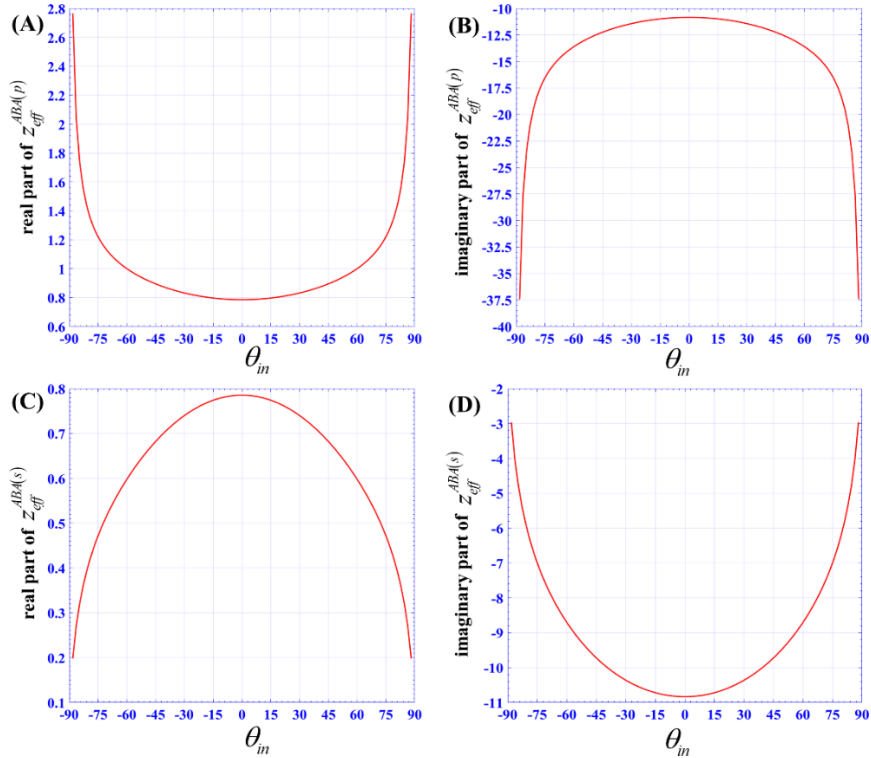

Fig. 9. (A) Real part of p-polarized effective impedance $z_{eff}^{ABA(p)}$ with respect to incidence angle $\theta_{in}$. (B) Imaginary part of p-polarized effective impedance $z_{eff}^{ABA(p)}$ with respect to incidence angle $\theta_{in}$. (C) Real part of s-polarized effective impedance $z_{eff}^{ABA(s)}$ with respect to incidence angle $\theta_{in}$. (D) Imaginary part of s-polarized effective impedance $z_{eff}^{ABA(s)}$ with respect to incidence angle $\theta_{in}$. $\mu_{r(a)}=1$, $\varepsilon_{r(a)}=5$, $\mu_{r(b)}=1$, $\varepsilon_{r(b)}=1$, and $d_a=5nm$, $d_b=10nm$.

As the same example in section 2.2, we take wavelength as 500 nm, $\mu_{r(a)}=1$, $\varepsilon_{r(a)}=5$, $\mu_{r(b)}=1$, $\varepsilon_{r(b)}=1$, and $d_a=5nm$  $d_b=10nm$. In this ABA structure case, the two materials A and B have no similar limitation as the AB periodic structure. From Eqs. (31a)-(31f), we can obtain the effective parameters $\theta_{eff}^{ABA}$, $\theta_{EGRA}^{ABA}$, $n_{eff}^{ABA(p,s)}$, and $z_{eff}^{ABA(t)(p,s)}$ with respect to incidence angle $\theta_{in}$.

Figs. 8(A), 8(B) and 8(C) are effective geometrical refractive angle $\theta_{EGRA}^{AB}$, p-polarized effective refractive index $n_{eff}^{ABA(p)}$, and s-polarized effective refractive index $n_{eff}^{ABA(s)}$ with respect to incidence angle $\theta_{in}$, respectively. Fig. 8(D) shows the breakdown of the Snell's law for EMT, where blue, red, and green dotted lines are $n_{in}\sin(\theta_{in})$, $n_{eff}^{ABA(p)}\sin(\theta_{eff}^{ABA(p)})$, and $n_{eff}^{ABA(s)}\sin(\theta_{eff}^{ABA(s)})$, respectively. Figs. 9(A), 9(B), 9(C) and 9(D) are real part of p-polarized effective impedance $z_{eff}^{ABA(p)}$, imaginary part of p-polarized effective impedance $z_{eff}^{ABA(p)}$, real part of s-polarized effective impedance $z_{eff}^{ABA(s)}$, and imaginary part of s-polarized effective impedance $z_{eff}^{ABA(s)}$ with respect to incidence angle $\theta_{in}$, respectively.

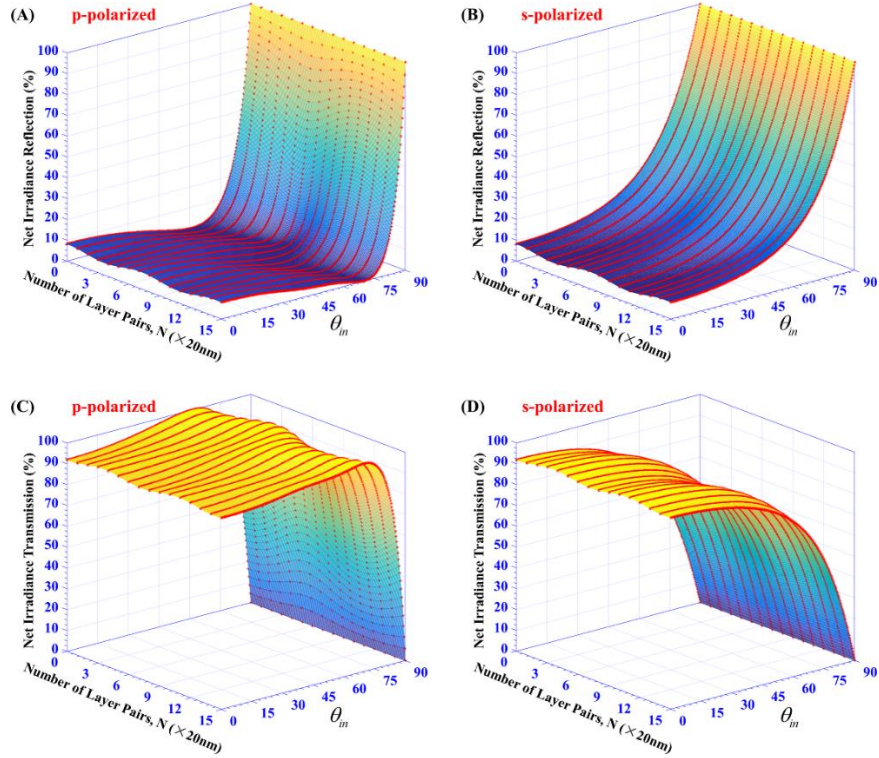

Fig. 10. (A) 3D plot of net irradiance reflection for p-polarized situation. (B) 3D plot of net irradiance reflection for s-polarized situation. (C) 3D plot of net irradiance transmission for p-polarized situation. (D) 3D plot of net irradiance transmission for s-polarized situation. ABA periodic structure with $\mu_{r(a)}=1$, $\varepsilon_{r(a)}=5$, $\mu_{r(b)}=1$, $\varepsilon_{r(b)}=1$, and $d_a=5nm$, $d_b=10nm$. $\varepsilon_{r(in)}=\mu_{r(in)}=\mu_{r(ou)}=1$, and $\varepsilon_{r(ou)}=3.25$.

Obviously, some effective parameters are infinite at the incidence angle of 90 degrees. This is because the material B has the same parameters as the vacuum, which is not consistent with the actual situation. If the parameter $\varepsilon_{r(b)}$ is larger, these infinite parameters become finite.

To verify no approximation of Eqs. (31a)-(31f), we also use TMT to simulate our ABA periodic structure and the corresponding EMT model for oblique incidence situation. We can obtain the effective parameters as the data in Figs. 8 and 9. Meanwhile, we take the relative permittivity and the relative permeability of the input port medium ($\varepsilon_{r(in)}$ and $\mu_{r(in)}$) as 1, the relative permeability of output port medium ($\mu_{r(ou)}$) as 1, and the relative permittivity of the output port medium ($\varepsilon_{r(ou)}$) as 3.25.

As shown in Fig. 10, the color surfaces are the EMT results, and the red asterisk dots are the TMT results. Figs. 10(A) and 10(C) are p-polarized results, and Figs. 10(B) and 10(D) are s-polarized results. For p-polarized light, there is zero reflection phenomenon, i.e. total transmission, at certain angles, which is connected to the Brewster angle. The red asterisk dots coincide perfectly with the color surfaces. Therefore, it reflect that there are no approximation between our EMT (i.e. Eqs. (31a)-(31f)) and TMT.

## 4. Conclusion

The concept of PMMs is important and interesting. We analyzed the transfer matrix of two kinds of 1D PMMs, and derived their effective parameters, for normal incidence and oblique incidence situations. There is no approximation of our EMT for these 1D PMMs. It's interesting and confusing that the effective refractive angle and effective refractive index don't obey the Snell's law, and the effective refractive angle is different from the effective geometrical refractive angle. It is challenging and fascinating to generate the 3D PMMs.

**Funding**

National Natural Science Foundation of China (No. 11775147); Guangdong Basic and Applied Basic Research Foundation (Nos. 2019A1515011474 and 2019A1515110130); Science and Technology Program of Shenzhen (Nos. JCYJ20200109105201936, JCYJ20190808115605501, and JCYJ20180305125443569).

**Disclosures**

The authors declare no conflicts of interest.

See Supplement 1 for supporting content.

**Reference**

1. T. J. Cui, D. R. Smith, and R. Liu, *Metamaterials: Theory, Design, and Applications* (Springer, 2010).
2. W. Cai and V. Shalaev, *Optical Metamaterials: Fundamentals and Applications* (Springer, 2010).
3. D. H. Werner and D.-H. Kwon, *Transformation Electromagnetics and Metamaterials: Fundamental Principles and Applications* (Springer, 2014).
4. A. Chipouline and F. Küppers, *Optical Metamaterials: Qualitative Models* (Springer, 2018).
5. P. Yeh, A. Yariv, and C.-S. Hong, "Electromagnetic propagation in periodic stratified media. I. General theory," J. Opt. Soc. Am. **67**(4), 423-438 (1977).
6. D. C. Flanders, "Submicrometer periodicity gratings as artificial anisotropic dielectrics," Appl. Phys. Lett. **42**, 492 (1983).
7. F. Xu, R.-C. Tyan, P.-C. Sun, Y. Fainman, C.-C. Cheng, and A. Scherer, "Fabrication, modeling, and characterization of form-birefringent nanostructures," Opt. Lett. **20**, 2457 (1995).
8. J. B. Pendry, "Negative refraction makes perfect lens," Phys. Rev. Lett. **85**, 3966 (2000).
9. D. R. Smith, W. J. Padilla, D. C. Vier, S. C. Nemat-Nasser, and S. Schultz, "Composite medium with simultaneously negative permeability and permittivity," Phys. Rev. Lett. **84**(18), 4184 (2000).
10. R. Shelby, D. Smith, and S. Schultz, "Experimental verification of a negative index of refraction," Science **292**, 77 (2001).
11. S. Zhang, W. Fan, N. C. Panoiu, K. J. Malloy, R. M. Osgood, and S. R. J. Brueck, "Experimental demonstration of near-infrared negative-index metamaterials," Phys. Rev. Lett. **95**, 137404 (2005).
12. V. M. Shalaev, W. Cai, U. K. Chettiar, H. K. Yuan, A. K. Sarychev, V. P. Drachev, and A. V. Kildishev, "Negative index of refraction in optical metamaterials," Opt. Lett. **30**, 3356 (2005).
13. X. Zhang, M. Davanço, Y. Urzhumov, G. Shvets, and S. R. Forrest, "From scattering parameters to Snell's law: a subwavelength near-infrared negative-index metamaterial," Phys. Rev. Lett. **101**, 267401 (2008).
14. B. Edwards, A. Alù, M. E. Young, M. Silveirinha, and N. Engheta, "Experimental verification of epsilon-near-zero metamaterial coupling and energy squeezing using a microwave waveguide," Phys. Rev. Lett. **100**, 033903 (2008).
15. R. Maas, J. Parsons, N. Engheta, and A. Polman, "Experimental realization of an epsilon-near-zero metamaterial at visible wavelengths," Nat. Photonics 7, 907 (2013).
16. H. Chen, B.-I. Wu, B. Zhang, and J. A. Kong, "Electromagnetic wave interactions with a metamaterial cloak," Phys. Rev. Lett. **99**(6), 063903 (2007).
17. N. Seddon and T. Bearpark, "Observation of the inverse Doppler effect," Science **302**(5650), 1537-1540 (2003).
18. P. Yu, L. V. Besteiro, Y. Huang, J. Wu, L. Fu, H. H. Tan, C. Jagadish, G. P. Wiederrecht, A. O. Govorov, and Z. Wang, "Broadband metamaterial absorbers," Adv. Opt. Mater. **7**(3), 1800995 (2019).

19. N. I. Landy, S. Sajuyigbe, J. J. Mock, D. R. Smith, and W. J. Padilla, "Perfect metamaterial absorber," Phys. Rev. Lett. **100**(20), 207402 (2008).
20. D. Wang, J. Song, J. Xian, Y. Tian, L. Chen, S. Ye, H. Niu, and J. Qu, "Characteristic analysis of broadband plasmonic emitting devices based on transformation optics," Opt. Express **23**(12), 16109 (2015).
21. D. Wang, J. Song, M. Xiong, G. Wang, X. Peng, and J. Qu, "Modified method for computing the optical force of the plasmonics nanoparticle from Maxwell stress tensor," J. Opt. Soc. Am. B **34**(1), 178 (2017).
22. D. Wang, J. Huang, Y. Lei, W. Fu, Y. Wang, P. Deng, H. Cai, and J. Liu, "Transparent conductive films based on quantum tunneling," Opt. Express **27**(10), 14344 (2019).
23. D. Wang, Y. Wang, J. Huang, W. Fu, Y. Lei, P. Deng, H. Cai, and J. Liu, "Low-cost and flexible anti-reflection films constructed from nano multi-layers of $TiO_2$ and $SiO_2$ for perovskite solar cells," IEEE Access **7**, 176394 (2019).
24. T. C. Choy, *Effective medium theory principles and applications* (Oxford University Press, 2016).
25. M. H. Cohen and J. Jortner, "Effective medium theory for the Hall Effect in disordered materials," Phys. Rev. Lett. **30**(15), 696-698 (1973).
26. D. Stroud, "Generalized effective-medium approach to the conductivity of an inhomogeneous material," Phys. Rev. B **12**(8), 3368-3373 (1975).
27. D. E. Aspnes, "Local-field effects and effective-medium theory: A microscopic perspective," Am. J. Phys. **50**(8), 704-709 (1982).
28. K. W. Jacobsen, J. K. Norskov, and M. J. Puska, "Interatomic interactions in the effective-medium theory," Phys. Rev. B **35**(14), 7423-7442 (1987).
29. X. C. Zeng, D. J. Bergman, P. M. Hui, and D. Stroud, "Effective-medium theory for weakly nonlinear composites," Phys. Rev. B, **38**(15), 10970-10973 (1988).
30. J. W. Haus, R. Inguva, and C. M. Bowden, "Effective-medium theory of nonlinear ellipsoidal composites," Phys. Rev. A **40**(10), 5729-5734 (1989).
31. D. R. Smith, S. Schultz, P. Markos, and C. M. Soukoulis, "Determination of effective permittivity and permeability of metamaterials from reflection and transmission coefficients," Phys. Rev. B **65**, 195104 (2002).
32. D.R. Smith and J. B. Pendry, "Homogenization of metamaterials by field averaging (invited paper)," J. Opt. Soc. Am. B **23**(3), 391-403 (2006).
33. T. Geng, S. Zhuang, J. Gao, and X. Yang, "Nonlocal effective medium approximation for metallic nanorod metamaterials," Phys. Rev. B **91**, 245128 (2015).
34. H.-Y. Xie, M.-Y. Ng, and Y.-C. Chang, "Analytical solutions to light scattering by plasmonic nanoparticles with nearly spherical shape and nonlocal effect," J. Opt. Soc. Am. A **27**(11), 2411-2422 (2010).
35. W. Yan, N. A. Mortensen, and M. Wubs, "Hyperbolic metamaterial lens with hydrodynamic nonlocal response," Optics Express **21**(12), 15026-15036 (2013).
36. J. Benedicto, R. Pollès, C. Ciracì, E. Centeno, D. R. Smith, and A. Moreau, "Numerical tool to take nonlocal effects into account in metallo-dielectric multilayers," J. Opt. Soc. Am. A **32**(8), 1581-1588 (2015).
37. J. Elser, V. A. Podolskiy, I. Salakhutdinov, and I. Avrutsky, "Nonlocal effects in effective-medium response of nanolayered metamaterials," Appl. Phys. Lett. **90**(19), 191109 (2007).
38. A. V. Chebykin, A. A. Orlov, C. R. Simovski, Yu. S. Kivshar, and P. A. Belov, "Nonlocal effective parameters of multilayered metal-dielectric metamaterials," Phys. Rev. B **86**(11), 115420 (2012).
39. L. Sun, J. Gao, and X. Yang, "Giant optical nonlocality near the Dirac point in metal-dielectric multilayer metamaterials," Opt. Express **21**(18), 21542 (2013).
40. R.-L. Chern, "Spatial dispersion and nonlocal effective permittivity for periodic layered metamaterials," Opt. Express **21**(14), 16514-16527 (2013).
41. B. Janaszek, M. Kieliszczyk, A. Tyszka-Zawadzka, and P. Szczepański, "Multiresonance response in hyperbolic metamaterials," Appl. Opt. **57**(9), 2135-2141 (2018).
42. Andrea Alù, "First-principles homogenization theory for periodic metamaterials," Phys. Rev. B **84**, 075153 (2011).
43. G. T. Papadakis, P. Yeh, and H. A. Atwater, "Retrieval of material parameters for uniaxial metamaterials," Phys. Rev. B **91**, 1554066 (2015).
44. S. Laroouche and V. Radisic, "Retrieval of all effective susceptibilities in nonlinear metamaterials," Phys. Rev. A **97**, 043863 (2018).
45. A. Rose, S. Laroouche, D. Huang, E. Poutrina, and D. Smith, "Nonlinear parameter retrieval from three- and four-wave mixing in metamaterials," Phys. Rev. E 82, 036608 (2010).
46. J. D. Jackson, *Classical Electrodynamics* (Wiley, 1999).
47. H. A. Macleod, *Thin-Film Optical Filters* (3rd ed., Institute of Physics, 2001).
48. A. V. Chebykin, A. A. Orlov, A. V. Vozianova, S. I. Maslovski, Y. S. Kivshar, and P. A. Belov, "Nonlocal effective medium model for multilayered metal-dielectric metamaterials," Phys. Rev. B **84**, 115438 (2011).
49. M. A. Gorlach and M. Lapine, "Boundary conditions for the effective-medium description of subwavelength multilayered structures," Phys. Rev. B **101**, 075127 (2020).
50. H. H. Sheinfux, I. Kaminer, Y. Plotnik, G. Bartal, and M. Segev, "Subwavelength multilayer dielectrics: ultrasensitive transmission and breakdown of effective-medium theory," Phys. Rev. Lett. **113**, 243901 (2014).
51. S. V. Zhukovsky, A. Andryieuski, O. Takayama, E. Shkondin, R. Malureanu, F. Jensen, and A. V. Larinenko, "Experimental demonstration of effective medium approximation breakdown in deeply subwavelength all-dielectric multilayers," Phys. Rev. Lett. **115**, 177402 (2015).

52. V. Popov, A. V. Lavrinenko, and A. Novisky, "Operator approach to effective medium theory to overcome a breakdown of Maxwell Garnett approximation," Phys. Rev. B **94**, 085428 (2016).
53. C. C. Lee and Y.-J. Chen, "Multilayer coatings monitoring using admittance diagram," Opt. Express **16**(9), 6119-6124, 2008.
54. Q.-Y. Cai, Y.-X. Zheng, D.-X. Zhang, W.-J. Lu, R.-J. Zhang, W. Lin, H.-B. Zhao, and L.-Y. Chen, "Application of image spectrometer to in situ infrared broadband optical monitoring for thin film deposition," Opt. Express **19**(14), 12969-12977, 2011.
55. K.Wu, C.-C. Lee, and T.-L. Ni, "Advanced broadband monitoring for thin lm deposition through equivalent optical admittance loci observation," Opt. Express **20**(4), 3883-3889, 2012.
56. Q. Y. Cai, H.-H. Luo, Y.-X. Zheng, and D.-Q. Liu, "Design of nopolarizing cut-off filters based on dielectric-metal-dielectric stacks," Opt. Express **21**(16), 19163-19172, 2013.
57. D. Wang, R. Zhou, Y. Wu, H. Cai, and Y. Zhang, "Improving external quantum efficiency by subwavelength nano multi-layered structures for opoelectronic devices," IEEE Access **8**, 189974-189981, 2020.